\documentclass[journal]{IEEEtran}

\usepackage{amsmath}
\usepackage{amssymb}
\usepackage{mathtools}
\usepackage{cases}
\usepackage{esvect}
\usepackage{graphicx}
\usepackage{balance}
\usepackage{subcaption}
\usepackage{lipsum}
\usepackage{multicol}
\usepackage{enumerate}
\usepackage{algorithm}
\usepackage[noend]{algpseudocode}
\usepackage{tikz}
\usepackage{booktabs}
\usepackage{xcolor}
\usetikzlibrary{shapes,arrows,chains}
\usetikzlibrary{shapes.geometric, arrows}
\usetikzlibrary{shapes,snakes}
\usepackage[latin1]{inputenc}
\usetikzlibrary{calc,patterns}
\usetikzlibrary{quotes,angles}
\usetikzlibrary{shapes,arrows}
\usetikzlibrary{automata, positioning}
\usetikzlibrary{arrows.meta}
\tikzset{%
  >={Latex[width=2mm,length=2mm]},
            base/.style = {rectangle, rounded corners, draw=black,
                           minimum width=4cm, minimum height=1cm,
                           text centered, font=\sffamily},
  activityStarts/.style = {base, fill=blue!30},
       startstop/.style = {base, fill=red!30},
    activityRuns/.style = {base, fill=green!30},
         process/.style = {base, minimum width=2.5cm, fill=orange!15},
}
\ifCLASSINFOpdf
\else
\fi
\begin{document}
\title{Metacognitive Decision Making Framework for Multi-UAV Target Search Without Communication}

\author{J. Senthilnath, \IEEEmembership{Senior~Member,~IEEE}, K. Harikumar \IEEEmembership{Member,~IEEE}
        and~S.~Suresh, \IEEEmembership{Senior~Member,~IEEE}
        
\thanks{\noindent J. Senthilnath is with the Institute for Infocomm Research, Agency for Science, Technology and Research (A*STAR), Singapore, 138632. e-mail: J\_Senthilnath@i2r.a$-$star.edu.sg}
\thanks{\noindent K. Harikumar is with the International Institute of Information Technology, Hyderabad-500032, India. e-mail: harikumar.k@iiit.ac.in}
\thanks{\noindent S. Suresh is  with the Department of Aerospace Engineering,
Indian Institute of Science, Bangalore 560012, India. e-mail:
vssuresh@iisc.ac.in}
  \thanks{\noindent This work has been submitted to the IEEE for possible publication. Copyright may be transferred without notice, after which this version may no longer be accessible.}
}
\markboth{Preprint version}%
{Shell \MakeLowercase{\textit{et al.}}: Bare Demo of IEEEtran.cls for IEEE Journals}
\maketitle
\begin{abstract}
This paper presents a Metacognitive Decision Making (MDM) framework inspired by human-like metacognitive principles. The MDM framework is incorporated in unmanned aerial vehicles (UAVs) deployed for decentralized stochastic search without communication for detecting and confirming stationary targets (fixed/sudden pop-up) and dynamic targets. The UAVs are equipped with multiple sensors (varying sensing capability) and search for targets in a largely unknown area. The MDM framework consists of a metacognitive component and a self-cognitive component. The metacognitive component helps to self-regulate the search with multiple sensors addressing the issues of "which-sensor-to-use", "when-to-switch-sensor", and "how-to-search". Based on the information gathered by sensors carried by each UAV, the self-cognitive component regulates different levels of stochastic search and switching levels for effective searching, where the lower levels of search aim to localize a target (detection) and the highest level of a search exploit a target (confirmation). The performance of the MDM framework with two sensors having a low accuracy for detection and increased accuracy to confirm targets is evaluated through Monte-Carlo simulations and compared with six multi-UAV stochastic search algorithms (three self-cognitive searches and three self and social-cognitive based searches). The results indicate that the MDM framework can efficiently detect and confirm targets in an unknown environment.
\end{abstract}
\graphicspath{{Images/}}
\begin{IEEEkeywords}
Decentralized search, Metacognitive, Uniform distribution, Brownian distribution, Levy distribution.
\end{IEEEkeywords}
\IEEEpeerreviewmaketitle

\section{Introduction}
\IEEEPARstart The Unmanned Aerial Vehicles (UAVs) are gaining popularity in applications like surveillance (vegetation analysis \cite{senpap1}, power-line monitoring \cite{senpap2}, smart cities \cite{pap3} and defense applications \cite{pap33}) and reconnaissance missions (crop pests \cite{pap4}, fire quenching \cite{pap5} and wildfire tracking \cite{smca1}) due to their easy usage in inhospitable areas. In most practical search and reconnaissance missions like disaster management, the location of targets is unknown. The targets can be distributed sparsely in a large and unknown area. The sudden appearance (pop-up) targets demands revisiting the search regions frequently \cite{popup}. In such scenarios, the reliability of the communication network to exchange information between UAVs decreases and hence increases the complexity of the problem further \cite{DGhose1}.  

Decentralized operation of UAVs with restricted communication or without communication reduces the delay and power requirement while searching a large area. A decentralized multi-UAV search is performed with or without a priori information about the target spatial distribution. When a priori information is available, the search technique can be optimized \cite{besada}. Some studies have formulated the optimization problem for the multi-UAV system based on team theory \cite{suresh}, game theory \cite{pap14}, and negotiation schemes \cite{pap16}. Another category of decentralized search algorithms creates a target probability map in real-time by discretizing the search space without any prior information \cite{poly, ant}; this approach needs an update at constant time intervals. Moreover, it generates a lower probability of revisiting a location where the target was not found at the last visit, hence not feasible while searching for sudden pop-ups and dynamic targets.

Identified stationary targets (fixed/sudden pop-up) and dynamic targets can be mitigated or tracked for information. Patrizi et al., \cite{Revnewref} proposed multi-target tracking and data collection framework for UAVs while addressing matching problems with targets. When there is no prior information about the target spatial distribution, the cognitive-based approach is more effective \cite{ppap17}. Since these approaches are solely on the sensor information, they are not computationally intensive (do not assign any target probability map), and hence can handle pop-up and dynamic targets efficiently. The cognitive-based approach consists of only self-cognitive or a combination of self and social-cognitive (communication among agents) elements. When a combination of self and social-cognitive elements is employed for search, the target is obtained based on the information from the individual sensor of a UAV \cite{smca2} and communication among UAVs \cite{ppso18, smca3}. The main drawback of the social-cognitive approach is to maintain communication among agents in large search spaces like forest regions or regions affected due to natural disasters. 

On the other hand, the self-cognitive search method doesn't rely on inter-UAV communication and is suitable for situations involving a large search region. The self-cognitive approach searches the target by navigating the agent with the waypoints in search space generated using various probability distributions (e.g. Levy, Uniform, Gaussian). The Levy Search (LS) is a stochastic search that uses Levy distribution to determine the search direction \cite{ppap17} and is useful when targets are distributed sparsely in a region \cite{ppap18}. Other popular stochastic searches are Brownian Search (BS) based on Gaussian distribution and Uniform-random Search (US) based on uniform distribution \cite{ghose}. Many decentralized algorithms were developed to imbibe UAVs with cognitive decision-making abilities, such as when-to-search and when-to-attack \cite{pap10,senpap3,pap11}.

Most of the existing cognitive-based approach performs a search using a single sensor. The use of multiple sensors for object detection and tracking for real-world applications like forest firefighting \cite{oms}, rescue operation during natural disasters \cite{sp1}, and precision agriculture \cite{sp2} is more effective than using a single sensor \cite{mm1, DGhose2, SingleMulti}. Unlike single sensor-based search techniques, the use of multiple sensors introduces the following challenging factors: a) "which among the multiple sensors to be used at a given time", b) "when the switching from one sensor to another should occur" and c) "how to self-regulate the stochastic search" to maximize the search efficiency. Hence, a generic methodology for multi-sensor based UAVs decentralized stochastic search to address these challenges. This study focuses on a novel metacognitive approach to adapt dynamic decision-making ability using multiple sensors. Recently, researchers in artificial intelligence have used the metacognition framework for designing robust and intelligent algorithms for pattern classification problems \cite{SateeshSuresh, KartickSuresh, DoraSuresh, Dhika1} and learning in a dynamic environment \cite{Josyula, WongSuresh} using the principle of metacognition. 

This paper addresses a Metacognitive Decision Making (MDM) framework for decentralized multi-UAV search based on the Nelson and Narens model, \cite{nelson90} of metacognition. The proposed decision-making system enables \emph{autonomy} in each UAV for self-regulating the strategies and \emph{self-awareness} based on the sensor information. In summary, the key contributions of our paper are three-fold:\\
1) The proposed MDM framework consists of two components: i) a self-cognitive component, which learns about the environment by sensing, and analyzing the information gathered from the sensors and acting upon it and, ii) a metacognitive component is a regulatory mechanism that monitors the effectiveness of the self-cognitive component that adapts the search mechanism in an unknown environment. Based on the monitoring signal, the metacognitive component regulates the search strategies and the choice of a sensor required to handle the environmental changes. Thus the individual's action and the information gained by them influence the metacognitive component so that their decision attains the overall goal efficiently.\\
2) The MDM framework-based multi-UAV stochastic search algorithm uses multiple sensors carried by each UAV for searching fixed, sudden pop-up and dynamic targets in the absence of inter-UAV communication. Each UAV performs a self-regulate search based on sensor information by switching between two or more stochastic techniques with different step lengths. A probabilistic comparison of the MDM framework with a single sensor-based stochastic search technique is provided to show the theoretical advantage of the proposed MDM.\\
3) The MDM framework search performance using two sensors is evaluated in a simulated setting and compared with various state-of-the-art stochastic search techniques employing a single sensor. For performance comparison, all the search techniques are evaluated in a decentralized setting without inter-UAV communication. A Monte-Carlo simulation was performed to analyze the efficiency, completeness, and scalability (in terms of the number of UAVs involved in the mission) of the proposed search technique. The comparison was also performed in the case of false detection of targets. The numerical results highlight the advantages of the proposed MDM framework for multi-UAV search compared to three state-of-the-art self-cognitive-based stochastic search techniques and three social-cognitive search techniques. Overall the MDM framework can achieve better search performance and completeness by adopting necessary stochastic techniques and sensors at different levels.

The rest of the paper is organized as follows: Section II presents the problem definition, Section III discusses the metacognitive decision making framework for multi-UAV search. The Monte-Carlo simulation results are discussed in section IV. The paper is concluded in section V.
\begin{table}[h]
\centering
\caption{List of symbols and their descriptions}
\label{ParaDesc}
\begin{tabular}{l l}
\hline
Symbols & Description \\ 
\hline
$U$, $T$, $S$ & UAVs, targets and sensors \\
$n_u$, $n_T$, $n$ & Total number of UAVs, targets and sensors \\
$n_i$, $n_p$ & Number of targets initially and pop-up targets \\
$X_T$, $X_U$ & Target position and UAV position ($m$)\\
$V_s$, $\kappa_s$ & Target speed ($m/s$) and sampling time ($s$)\\
$V_U$, $Q_U$ & UAV velocity and UAV velocity reference ($m/s$) \\
$D_{ks}$, $D_{kt}$ & Signal strength of  $k^{th}$ detection sensor\\
                   & and detection threshold \\
$C_{ns}$, $C_{nt}$ & Signal strength of the confirmation sensor\\ 
                   & and confirmation threshold \\
$r_k$, $r_n$ & Sensing radius of detection and confirmation sensor ($m$) \\
$T_{nc}$ & Number of targets confirmed in a fixed time duration\\
$\mu(T_{nc})$ & Expected value of the number of targets confirmed \\
$N_f$ & Number of finite-time missions executed by the UAVs \\
$T_S$ & time taken to confirm all the targets ($s$) \\
$\mu(T_{S})$ & Expected value of time taken to confirm all targets ($s$)\\
$N_T$ & Number of multi-UAV missions corresponding to \\
      & search completeness \\
$P_q$ & $q^{th}$ waypoint of a UAV\\
$\delta$ & Search step length ($m$)\\
$D(S_t)$ & Direction of motion generated from stochastic process $S_t$ \\
$\alpha$, $\lambda$ & Parameters of Levy distribution\\
$\sigma$ & Parameter of Brownian distribution \\
$v_u$, $v_l$ & Parameters of Uniform distribution\\
$C_I$ & Confirmation index \\
\hline
\end{tabular}
\end{table}

\section{Multi-Sensor UAVs Search Problem}
\begin{figure}
    \centering
    \includegraphics[height=6.5cm, width=7.5cm]{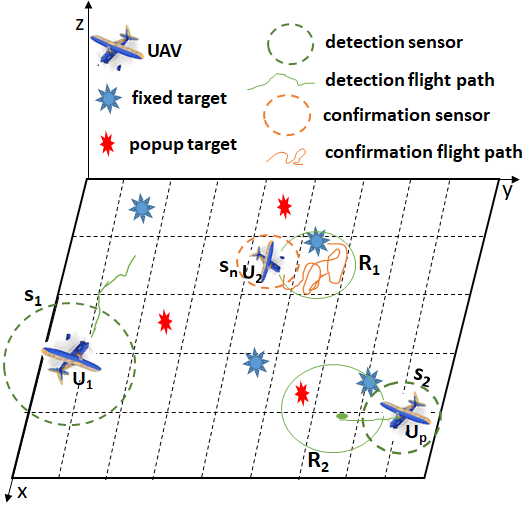}
    \caption{Illustration of multi-UAV search with different sensors}
    \label{scalefig}
\end{figure}

Table \ref{ParaDesc} provides the list of symbols used in this work and their description. In a typical scenario, the detection and confirmation of targets by UAVs is challenging due to: i) UAVs are searching in a large area, ii) no communication between UAVs, and iii) no prior information about the target spatial distribution is available. In this study, we assume the targets are fixed, sudden pop-ups and dynamic. The UAVs perform a decentralized search for detecting and confirming the targets without communicating between them. The objective of UAVs' mission is to search and detect the fixed, sudden pop-up and dynamic targets distributed in a large area. A generic scenario of using UAVs carrying multiple sensors for fixed or pop-up targets detection and confirmation is shown in Fig. \ref{scalefig}. The region in which UAVs operate is denoted by \textbf{R} $\in$ $\mathbb{R}^3$. UAVs are denoted by $U_1,\,U_2,\,...,U_{n_u}$ where $n_u$ is the total number of UAVs.  Each UAV carries $n_s$ number of sensors denoted by  $S_1,\,S_2,\,...,S_{n}$. The targets are denoted by $T_1,\,T_2,\,...,T_{n_T}$, where $n_T$ is the total number of targets. Let the time $t$=0 denote the start of the mission. Here, fixed, sudden pop-up and dynamic targets are considered. Initially, fixed targets are present in the region from the start of the mission ($t$=0), whereas the sudden pop-up targets appear at any time $t>0$ during the mission. Let the number of targets initially present be denoted by $n_{i}$, and the number of pop-up targets at any time $t>0$ is denoted by $n_p$. The number of pop-up targets is mathematically expressed as,
\begin{equation}
    n_p= \sum_{i}^{} \Delta(t-t_i)
\end{equation}
where $t_i>0$ is the time at which the $i^{th}$ pop-up target appeared in the search region and the unit step function 
\begin{equation}
\Delta(t-t_i)=\left\{
                \begin{array}{ll}
                  1, \,\, \forall\, t\geq t_i\\
                  0, \,\, \forall\, t < t_i
                \end{array}
              \right.
\label{scon2}
\end{equation}
Therefore, $n_T=n_{i}+n_p$. 

For dynamic targets, the position coordinate of $i^{th}$ target at time $t$ is given by $X_{T_{i}}(t)$ $\in$ $\mathbb{R}^2$. The motion model of the dynamic target is modelled by random walk as given below,
\begin{equation}
    {X}_{T_i}(k+1)={X}_{T_i}(k) + R(a,b) * V_s * \kappa_s
\end{equation}
where $R(a,b)$ denotes the random number generated from a uniform distribution within an interval $(a,b)$, $V_s$ denotes the constant speed in $m/s$ and $\kappa_s$ is the sampling time where $t = k*\kappa_s$ and $k$ is the discrete-time instant.

The UAVs fly at a constant altitude above the ground level, hence the altitude is not considered while defining its position coordinates. The position coordinate of the $i^{th}$ UAV at a time $t$ is denoted by $X_{U_i}(t) \in \mathbb{R}^2$. Similarly, for the $j^{th}$ target the position coordinate is denoted by $X_{T_j} \in \mathbb{R}^2$. The UAVs fly from one waypoint to another using velocity control. The dynamics of $i^{th}$ UAV are given by,
\begin{equation}
    \dot{X}_{U_i}(t)={V}_{U_i}(t)
\end{equation}
\begin{equation}
  \dot{V}_{U_i}(t)=-\frac{1}{\tau}{V}_{U_i}(t)+\frac{1}{\tau}{Q}_{U_i}(t)  \end{equation}
where ${V}_{U_i}(t)$ and ${Q}_{U_i}(t)$ denote the velocity coordinate and the velocity reference respectively for the UAV and $\tau>0$ is the time constant for the first-order closed-loop velocity control loop.
The following assumptions are made in this study regarding UAV flight.\\
\textit{Assumption I}: each UAV is capable of performing autonomous waypoint navigation in the presence of atmospheric disturbances.\\
\textit{Assumption II}: all UAVs are flying at the same speed denoted by $V_U$.\\
\textit{Assumption III}: UAVs equipped with inter-UAV collision and obstacle avoidance strategy. \\
The assumptions mentioned above are not conservative, as recently, advanced control techniques are applied to UAVs for wind disturbance rejection and to maintain steady speed \cite{xiao}. For the case of \textit{Assumption III}, a simple strategy would be that each UAV flies at an altitude greater than the obstacle height and also different from other UAVs.\\
The following assumptions are made regarding the search environment and targets.\\
\textit{Assumption IV}: the search environment is unknown with no prior information available regarding the target spatial distribution.\\
\textit{Assumption V}: The targets are stationary, but they can suddenly pop-up at a location. Hence the environment is dynamic with respect to the number of targets at a given time. Moreover, the targets are identical with the same sensing signature.

In Fig. \ref{scalefig}, UAV $U_1$ is searching the entire region \textbf{R} to detect the target using the sensor $S_1$. Whereas, the UAV $U_p$ is searching in a region $R_2 \subset$ \textbf{R} using the sensor $S_2$ to detect the target. The first $n-1$ sensors are used for target detection and the $n^{th}$ sensor is used for target confirmation. The UAV $U_2$ is searching in a region $R_1 \subset$ \textbf{R} for target confirmation using the sensor $S_{n}$. Each UAV undergoes a multi-sensor search and uses one sensor at a time for either target detection or target confirmation. The flight path generated during the detection or confirmation phase is stochastic in nature, governed by different probability density functions. To facilitate the detection and subsequent confirmation of the target, each UAV is equipped with multiple sensors. The sensor model adopted in this study is explained below.

\subsection{Detection and Confirmation Sensor}
The detection and confirmation sensors carried by a UAV capture the signal with strength proportional to the target's distance. These sensors are common in detecting targets like fire (sensing temperature) and mobile phones (radio frequency signal) based on signal strength. Generally, the signal strength increases as the UAVs approach the target. Let the strength of the signal captured by $k^{th}$ detection sensor $S_k$ at a time $t$ be denoted by $D_{ks}(t)$. If the value of $D_{ks}(t)$ exceeds the detection threshold denoted by $D_{kt}$, then a target is considered as detected. The sensing range of the detection sensor is considered homogeneous in all directions. Considering a case where the $i^{th}$ UAV is searching using the detection sensor $S_k$ having a sensing radius $r_k$, the following condition implies that a target is detected and the search region is further localized.
\begin{equation}
  D_{ks}(t)> D_{kt}   \Rightarrow ||X_{U_i}(t)-X_{T_j}||_2<r_k
   \label{detect}
\end{equation}
Similarly, the $j^{th}$ target is said to be confirmed by the $i^{th}$ UAV using a confirmation sensor denoted by $S_{n}$ if the following condition is satisfied.
\begin{equation}
   C_{ns}(t)> C_{nt}   \Rightarrow ||X_{U_i}(t)-X_{T_j}||_2<r_{n}
   \label{confirm}
\end{equation}
where $C_{ns}(t)$ is the strength of the signal received by the confirmation sensor and $C_{nt}$ is the confirmation threshold. The sensing radius of the confirmation sensor is denoted by $r_n$. The detection sensor can only give the spatial distribution of the target, whereas the confirmation sensor is used to obtain the exact target location. It is useful in scenarios like looking for a person (target) lost in a natural disaster. The cellular phone signal detector can be used to estimate the spatial distribution of the target, and a camera can be used to confirm the target. Here the confirmation threshold can be the minimum size of the target estimated from camera images. The detection sensor can result in a false positive, whereas the confirmation sensor always yields a true positive or true negative.

Two metrics are used to quantify the performance of the mission, considering the fact that it could be of either fixed-time duration or continued till all the targets in the region are confirmed. The number of targets confirmed ($T_{nc}$) during a mission of fixed time duration reflects the time efficiency of the search algorithm. For finite-time missions, the performance is quantified in terms of the expected value of the number of targets confirmed given by,
\begin{equation}
    \mu(T_{nc})=\frac{1}{N_f} \sum_{i=1}^{N_f} T_{nc}
\end{equation}
where $N_f$ is the number of finite-time missions executed by the UAVs. The mission continues until all the targets are confirmed, the time taken to confirm all the targets ($T_S$) needs to be minimal. This is related to the completeness of the search algorithm, and the performance is quantified by the expected value of time taken to confirm all targets given by,
\begin{equation}
      \mu(T_{S})=\frac{1}{N_T} \sum_{i=1}^{N_T} T_{S}  
\end{equation}
where $N_T$ is the number of missions executed by the UAVs to confirm all the targets. The stochastic search technique adopted along with the sensor to be used by the UAV during each level of search is the decision variable. The solution to the above-mentioned problem is addressed in a metacognitive decision making (MDM) framework discussed in the next section.

\section{MDM framework for multi-UAV search}
In this section, we present the proposed MDM framework for decentralized multi-UAV search with multiple sensors. The section concludes with the probabilistic comparison of the MDM-based search with the self-cognitive stochastic search.

\begin{figure*}
    \centering
    \includegraphics[height=6cm, width=16cm]{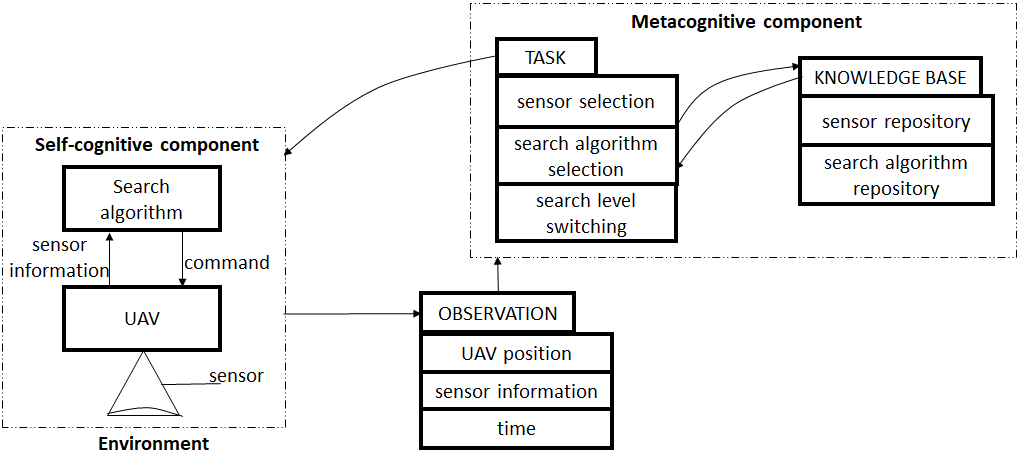}
    \caption{MDM framework for decentralized UAV search. Here, both self-cognitive and meta-cognitive are implemented within the same UAV. Self-cognitive consists of a search algorithm that is currently in use, navigation command, sensor module and sensor information. Meta-cognitive combines knowledge base and task allocation, i) the knowledge base consists of a collection of search algorithms, and a collection of sensors and their sensing characteristics, ii) task allocation determines the selection of the sensor and search algorithm that is being used in the self-cognition part. If there is only a single sensor and single search algorithm then only self-cognition is adapted (basically meta-cognition boils down to self-cognition as there is no selection of sensor or search algorithm). Whereas meta-cognition plays a role when we need to select one among multiple search algorithms and one among multiple sensors dynamically.}
    \label{mdmf}
\end{figure*}

The block diagram of the proposed MDM framework for each UAV is shown in Fig. \ref{mdmf}. The framework contains a self-cognitive component, a metacognitive component and observation. The metacognitive component regulates the self-cognitive based on observation. The self-cognitive component  is comprised of the following elements:  
\begin{itemize}
\item \textit{Search algorithm}: generates waypoints in the search space, randomly drawn from different stochastic processes based on the sensor information. Inspired by nature, the random displacement taken by a UAV when there is no communication between them derived from either Levy distribution ($LD$), Brownian distribution ($BD$) or Uniform distribution ($UD$) \cite{ppap18,stigpap,pap19,pap20}. The $q^{th}$ waypoint $(P_{q})$ and $(q+1)^{th}$ waypoint $(P_{q+1})$ of a UAV is related by,
\begin{equation}
    P_{q+1}=P_{q}+\delta*D(S_t)
    \label{Eq7}
\end{equation}
\noindent where $\delta$ is the step length and $D(S_t)$ is the direction of motion generated by any of the above-mentioned stochastic processes, ie. $S_t \in \{LD,\,BD,\,UD \}$. Let the two samples generated from $S_t$ be $\bar{S}$=($s_x$,$s_y$) then the direction $D(S_t)$ is along  ($\frac{s_x}{\parallel \bar{S} \parallel_2}$, $\frac{s_y}{\parallel \bar{S} \parallel_2}$) from the current location.
\item \textit{Navigation command}: generates the velocity/heading commands for the inner loop control system of the UAV based on the waypoints generated by the \textit{search algorithm}.
\item \textit{Sensor module}: is the physical device used for searching the targets. At any given time, either one of the detection sensors or the confirmation sensor is in use. 
\item  \textit{Sensor information}: is the strength of the signal received by the detection sensor or the confirmation sensor as given in (\ref{detect}) and (\ref{confirm}) respectively. The information collected from the sensor module is utilized to detect/confirm a target.
\end{itemize}
The observation component consists of the time information $t$, position of the $i^{th}$ UAV $X_{U_i}(t)$ and the sensor information $D_{ks}(t)$ or $C_{ks}(t)$. The observation component is the input to the metacognitive component.

The metacognitive component is composed of two main elements: \textit{knowledge base} and \textit{task allocation}, and their sub-elements.

\begin{itemize}
 \item \textit{Knowledge base}: contains the sub-elements, \textit{sensor repository} and \textit{search algorithm repository}. The \textit{knowledge base} is updated when a new sensor/search technique is added or the performance details of a search technique are obtained from analysis or through experiments.

\begin{itemize}
  \item \textit{Sensor repository}: contains the category (detection/confirmation) of the sensors carried by a UAV and their sensing radius ($r_1,\,r_2,\,...,r_{n}$).

\item \textit{Search algorithm repository}: contains the collection of stochastic search techniques adopted by a UAV. Here the repository contains the Levy search ($LS$), Brownian search ($BS$) and Uniform-random search ($US$). The Levy search ($LS$) is a  stochastic search whose search direction for each UAV is drawn from the Levy distribution. The probability density ($f(.)$) of a random variable $v$ under Levy stochastic process is given by \cite{levy1},
 \begin{equation}
   f(v)= \frac{1}{\pi} \int_{0}^{\infty} e^{(-\alpha m^\lambda)}\, cos(vm) \,\, dm
   \label{levy}
 \end{equation}
where $\lambda$ is an index satisfying 0$<\lambda \leq$2 and $\alpha$ is a scaling parameter (normally set to 1). For Levy search, $D(S_t) \sim f(v)$ in (\ref{levy}), where $\sim$ denotes the random number drawn from the distribution. Since a closed-form expression for the integral in (\ref{levy}) cannot be obtained, random numbers that follow Levy stochastic process are generated using the algorithm given in \cite{levy1}.

The Brownian search ($BS$) essentially uses a Gaussian distribution with zero mean and variance $\sigma$ given by,
\begin{equation}
    D(S_t) \sim N(0,\sigma) 
    \label{Eq9}
\end{equation}
\noindent where $\sim$ means the samples should be drawn using the probability density given by, 
\begin{equation}
    f(v) = \frac{1}{\sqrt{2\pi \sigma^2}} e^{(-{\frac{v^{2}}{2\sigma^2}})}
    \label{Eq10}
\end{equation}

When the uniform distribution is used, the resulting uniform-random search ($US$) draws samples from the uniform distribution with probability density given by,
\begin{equation}
    f(v)=\frac{1}{v_u-v_l}
\end{equation}
for $v \in [v_l,\,v_u]$ and $f(v)=0$ otherwise. Here $v_l$ and $v_u$ are the lower and upper bound respectively on the value taken by the random variable $v$.
\end{itemize}

 \item \textit{Task allocation}: decides the level of search operation and the corresponding sensor and the stochastic search technique employed. As there is $n_s$ number of sensors, there are $n_s$ levels of search. \textit{Task allocation} contains three  sub-elements,  \textit{sensor selection},  \textit{search algorithm selection} and   \textit{search level switching}. \textit{Task allocation} takes input from \textit{knowledge base} and \textit{observation} block to make the decisions.
 
\begin{itemize}
 \item \textit{Sensor selection}: decides the sensor to be used for search by the UAV. The UAV retains the current sensor or switches to a different one. Initially, the search starts with the sensor $S_1$ having the highest sensing range denoted by $r_1$. When the sensor information corresponds to the condition given in (\ref{detect}), then the search area is localized within a circle of radius $r_1$, with the current UAV position as the center of the circle. To further localize the search space, the sensor $S_2$ with the second-highest sensing range denoted by $r_2$ is used. This process continues until the search reaches the confirmation level where the sensor $S_{n_s}$ with the highest accuracy and the lowest range denoted by $r_{n_s}$ is used to confirm the target location. 

\item \textit{Search algorithm selection}: decides on selecting the appropriate search technique $S_t \in \{LD,\,BD,\,UD \}$ and the step length  $\delta \in \{\delta_1,\,\delta_2,\,...,\delta_{n-1},\, \delta_{n} \}$. Here $\delta_1$ is the step length when sensor $S_1$ is employed and similarly continues till $\delta_{n}$ is the step length when the confirmation sensor $S_{n}$ is employed. The following relation holds for the step length, $\delta_1 > \delta_2>....>\delta_{n-1}> \delta_n$. As each level of search localize the search space with respect to the target location, the condition $\delta_2 \leq r_1$, $\delta_3 \leq r_2$, ..., $\delta_{n} \leq r_{n-1}$ is employed.

\item \textit{Search level switching}: it is done when the condition given in (\ref{detect}) or (\ref{confirm}) is satisfied. The first $n-1$ level corresponds to target detection and the $n^{th}$ level corresponds to the target confirmation. To prevent a UAV to undergo an indefinite search at a particular level, without confirming a target, the following step is adopted. When a UAV enters the $n^{th}$ level of search, a confirmation index ($C_I$) is set to a value closer to one denoted by $C_{Imax}$. As the exploitation of the specified search region progress without successful confirmation of the target, the value of $C_I$ is reduced as given below.
\begin{equation}
    C_I=\Gamma+(C_{Imax}-C_{Imin})e^{(\frac{-t_s}{\kappa})}
\end{equation}
where $C_{Imin}$ is the lower bound for the confirmation index, $0<\Gamma<C_{Imin}$ is a constant, $t_s$ is the time spent on the $n^{th}$ level of search and $\kappa>0$ is a constant. If the condition,
\begin{equation}
  C_I<C_{Imin}
  \label{cond1}
\end{equation}
is satisfied, then the $n^{th}$ level of search is terminated and the first level is resumed. The same approach is used in a search level $k$, where  $k \in \{2,\,3,\,n-1 \}$, in order to prevent a UAV from indefinitely searching a region without further localizing the target, where the confirmation index ($C_I$) replaced by the detection index (${D^k}_I$). The condition analogous to (\ref{cond1}) for the $k^{th}$ detection level is given by,
\begin{equation}
  {D^k}_I<{D^K}_{Imin}
  \label{cond2}
\end{equation}
where ${D^K}_{Imin}$ is the lower bound for the detection index for the $k^{th}$ detection level.
\end{itemize}
\end{itemize}

 \begin{figure*}
    \centering
    \includegraphics[height=8cm, width=17cm]{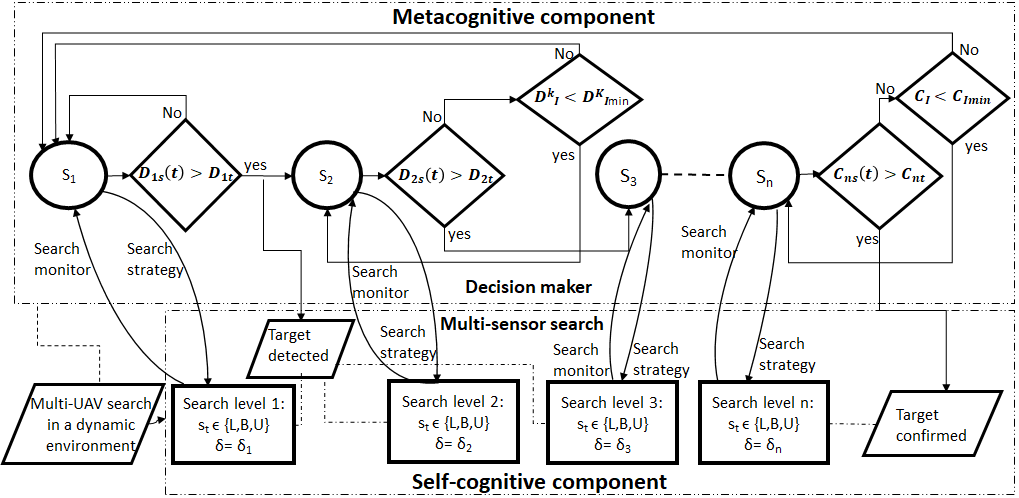}
    \caption{Flow diagram of the metacognitive decision making framework for UAVs search.}
    \label{scalefiga2}
\end{figure*}

The flow diagram of the MDM framework is shown in Fig. \ref{scalefiga2}. Each UAV starts with the first level of search where $S_t \in \{LD,\,BD,\,UD\}$, $\delta=\delta_1$ and a sensor with sensing radius $r_1$, till it detects a target according to the condition stated in (\ref{detect}) and then switch over to the second level to further localize the search space. The initial detection leaves a smaller search region for the second level of the search than the first level of the search. The second level uses a sensor with sensing radius $r_2$, $S_t \in \{LD,\,BD,\,UD \}$ and $\delta=\delta_2$. This process continues until the confirmation level is reached, where a sensor with sensing radius $r_n$, $S_t \in \{LD,\,BD,\,UD\}$ and $\delta=\delta_n$ is used to confirm the target. The $k^{th}$ level of search explores the search area with respect to the $(k+1)^{th}$ level of search, whereas the $(k+1)^{th}$ level of search exploits the search area with respect to $k^{th}$ level of search.  When the condition given in (\ref{confirm}) is satisfied, then a target and its location is confirmed, and then the UAV switch back to the first level of search for exploring the search space to detect other unknown targets. During each level of search, if the condition given in (\ref{cond1}) or (\ref{cond2}) is satisfied, then the UAV switch to the first level of search. 

\subsection{Probabilistic comparison of the MDM framework and a single sensor-based stochastic search}
For any self-cognitive based continuous random search with a single sensor, the probability of confirming a target by a UAV ($P_{CS}(t)$) at a time $t$ is given by \cite{ps},
\begin{equation}
    P_{CS}(t)=1-e^{(\frac{-2r_n ||V_{U}||_2 t}{A})}
    \label{p1}
\end{equation}
where $||V_{U}||_2$ is the speed of the UAV and $A$ is the total search area.

The probability of confirming a target ($P_{CM}(t)$) by a UAV for the case of MDM framework-based search (provided that the detection for $(n-1)^{th}$ level has happened at a time $t_d<t$) is given by,
\begin{equation}
     P_{CM}(t)=1-(1-P_{CS}(t_d))(e^{(\frac{-2r_n ||V_{U}||_2 (t-t_d)}{A_{n-1}})})
    \label{p3}
\end{equation}
where $A_{n-1}=\pi {r_{n-1}}^2$ is the sensing area of the $(n-1)^{th}$ detection sensor. Till $t \leq t_d$, the probability of confirming a target is the same for both cases, ie.
\begin{equation}
P_{CM}(t)=P_{CS}(t),\, \forall  t<t_d  
\end{equation}
 
\noindent It can be verified using (\ref{p3}) that the condition $P_{CM}(t)>P_{CS}(t)$ occurs when
\begin{equation}
    \frac{t-t_d}{A_{n-1}}>\frac{t-t_d}{A}
\end{equation}
The above condition is valid $\forall t \geq t_d$, since $A_{n-1}<A$. The above analysis indicates that for a given time instant, the probability of confirming a target for the proposed stochastic search is greater than or equal to any of the single sensor-based stochastic search techniques, ie.
\begin{equation}
  P_{CM}(t) \geq P_{CS}(t),\, \forall t>0 
\end{equation}

\section{Performance Evaluation}
In this section, numerical simulation results for the proposed MDM framework are presented initially, followed by a comparison with self and social-cognitive based search techniques. For the proposed MDM framework, we consider the case where $n=2$, i.e. each UAV is equipped with a single detection sensor and a confirmation sensor. The self and social-cognitive based search use only one sensor, so they have been categorized into single sensor-based search (SSS) techniques. In the case of social-cognitive based search, UAVs communicate with each other to share the search information obtained through self-cognition. The region under which UAVs perform search operations is of dimension L $\times$ W = 20 $km$ $\times$ 20 $km$. Each UAV follows the waypoints in the search space generated by the search technique. The conventional inner loop and outer loop control architecture is assumed here for the UAV \cite{beard}. The velocity commands required for autonomous waypoint navigation are generated using pursuit guidance law \cite{hari1}. The parameters and their values used in the simulation are given in Table \ref{param}.
\begin{table}[h]
\centering
\caption{Parameters and their values used in simulation}
\label{param}
\begin{tabular}{l l}
\hline
Parameter& Value              \\ 
\hline
LS $\lambda$, $\alpha$ & 1, 1   \\
BS $\sigma$ & 1   \\
US $v_u$, $v_l$ & 1, 0   \\
Search region, L $\times$ W& 20,000 $m$ $\times$ 20,000 $m$           \\
Number of UAVs, $n_u$ & 12\\
UAV speed, $||V_U||_2$& 20 $m/s$           \\
UAV time constant, $\tau$& 0.33 $s$          \\ 
Number of sensors, $n$ &  2 \\
Detection sensor radius, $r_1$ & 10$\%$ of L ($m$)             \\
Confirmation sensor radius, $r_2$ & 2$\%$ of L ($m$)   \\         Step length $\delta_1$, $\delta_2$& 20$\%$ of L ($m$), 10$\%$ of L ($m$) \\
Number of fixed targets, $n_i$ & 10 \\
Number of pop-up targets, $n_p$ & 10 (maximum) \\
Number of dynamic targets, $n_d$ & 10 \\
\hline
\end{tabular}
\end{table}

The simulations are carried out in MATLAB software installed in the Windows-10 operating system, with Intel-R Core-i7, 1.8 GHz processor and 16 GB RAM. Twelve UAVs were considered for performing the search mission with starting location distributed along the periphery of the search area. Each UAV is flying at a constant speed, $||V_U||_2\,=\,20\, m/s$. Three separate categories of simulations were carried out: i) initially fixed targets, ii)  sudden pop-up targets, and iii) dynamic targets. The video corresponding to the numerical simulation performed  for some of the scenarios in MATLAB (R2019b) can be found in the web link\footnote{https://youtu.be/THpaxv-1iDo}.\\
\emph{i) Fixed targets:} Here ten targets are placed in the search area at locations, $T_1$=(2$km$, 2$km$), $T_2$=(4$km$, 16$km$), $T_3$=(10$km$, 10$km$), $T_4$=(16$km$, 12$km$), $T_5$=(18$km$, 2$km$), $T_6$=(3$km$, 10$km$), $T_7$=(5$km$, 4$km$), $T_8$=(13$km$, 8$km$), $T_9$=(15$km$, 5$km$), $T_{10}$=(18$km$, 11$km$) respectively, where $n_i$=10, $n_p$=0 and $n_T$=$n_i$ $\forall t \geq 0$.  \\
\emph{ii) Pop-up targets:} The target locations are the same for the case of initially fixed targets and sudden pop-up targets. Each target suddenly appears at a time interval of 200 seconds with $t_i=200i$, where $n_i$=0 and $n_p$=$n_T$.\\
\emph{iii) Dynamic targets:} Here we consider target speed $v_s$ as 5 $m/s$, random number $R(a,b)$ is generated within the range a=-0.2 and b=+1 and sampling time $\kappa_{s}$=0.1 seconds.

The performance is evaluated in terms of a) search efficiency and b) search completeness. The search efficiency is evaluated by the number of targets confirmed ($T_{nc}$) while searching for a fixed time duration. The fixed time duration for the case of both initially fixed targets and dynamic targets is set to  1000 seconds, and for the case of sudden pop-up targets, it is set to 2000 seconds. The mean and standard deviation of the number of targets confirmed ($\mu(T_{nc})\pm\sigma(T_{nc})$) is computed for 1000 Monte-Carlo runs. The second performance criterion is the time taken ($T_S$) to search and confirm all the ten targets. The mean and standard deviation  ($\mu(T_S)\pm\sigma(T_S)$) for 1000 Monte-Carlo runs are computed for all the search techniques for evaluating the completeness of the search mission. For completeness of the search, it is desirable to have a lower mean and standard deviation for the time taken. 

\subsection{MDM framework-based search}
The numerical simulation results for MDM framework are presented in this section. The search performance varies depending on the step length used in the first level of search ($\delta_1$) and the probability distribution used for waypoint generation in the two levels of search.

\noindent \textit{$A_1$: Effect of step length ($\delta_1$) in the MDM search} \\
The effect of variation in step length of the first level of search ($\delta_1$) of the MDM framework on the search performance for fixed targets, as shown in Table \ref{simtable3}. The step length for the second level of search is set to the sensing radius of the detection sensor as given in Table \ref{param}, i.e. $\delta_2=r_1$. The step length is varied from 10$\%$ of $L$  to 50$\%$ of $L$, and Monte-Carlo simulations are performed for 1000 runs. The performance of the search initially increases with an increase in step length but later decreases. A step length of 20$\%$ of L gives the best search performance. For all three cases, the step length of 20$\%$ of L for the first level yields the best search results, as shown in Table \ref{simtable3}. For all three cases, LS is used in the first level, and BS is used in the second level of the search.

\begin{table}[h]
\centering
\caption{Effect of variation in step length on the MDM search with LS and BS for 10 targets/12 UAVs}
\label{simtable3}
\begin{tabular}{  c  c  c }
\hline
Step Length &  $\mu(T_{nc})\pm\sigma(T_{nc})$       &   $\mu(T_S)\pm\sigma(T_S)$     \\ 
$\delta_1$ & & (seconds)\\
        \hline
        Fixed targets &($n_i$=10 and $n_u$=12)&\\
\hline
10$\%$ of L          &  3.954$\pm$1.399  &  5532$\pm$2880 \\
20$\%$ of L           &  \textbf{4.658}$\pm$1.396  &  \textbf{4001}$\pm$1834 \\
30$\%$ of L          &  4.370$\pm$1.385  &  5640$\pm$3197 \\
40$\%$ of L          &  4.135$\pm$1.412  &  6216$\pm$3568 \\
50$\%$ of L          &  4.094$\pm$1.419  &  5721$\pm$3002 \\
\hline
Pop-up targets &($n_i$=0, max($n_p$)=10 and $n_u$=12)&\\
    \hline
10$\%$ of L          &  4.375$\pm$1.315  &  6606$\pm$2807 \\
20$\%$ of L           &  \textbf{5.106}$\pm$1.348  &  \textbf{4996}$\pm$1800 \\
30$\%$ of L          &  4.772$\pm$1.307  &  5726$\pm$2395 \\
40$\%$ of L          &  4.667$\pm$1.354  &  6409$\pm$2832 \\
50$\%$ of L          &  4.531$\pm$1.316  &  6753$\pm$3130 \\
\hline
Dynamic targets & ($n_d$=10 and $n_u$=12) &\\
    \hline
10$\%$ of L & 3.604$\pm$1.404 & 2772$\pm$1102 \\
20$\%$ of L & \textbf{4.424}$\pm$1.499 & \textbf{2382}$\pm$893 \\
30$\%$ of L & 4.265$\pm$1.443 & 2619$\pm$835 \\
40$\%$ of L & 3.819$\pm$1.377 & 2856$\pm$916 \\
50$\%$ of L & 3.739$\pm$1.473 & 2676$\pm$993 \\
\hline
\end{tabular}
\end{table}

\noindent \textit{$A_2$: Effect of different probability distribution at different levels of the MDM search.}\\
The nature of the probability distribution used in level 1 and level 2 of the MDM search affects the search performance. Monte-Carlo study, simulations are performed 1000 times to analyze the effect of the distribution used in level 1 and level 2 of the search. The results are shown in Table \ref{simtable4} for the case of 10 fixed targets where the Levy distribution is the best for level 1 search. For the level 2 search, the best value of $\mu(T_{nc})$ is for the Levy distribution, and the best value of $\sigma(T_{nc})$ is for the Brownian distribution. Similarly, the best value of $\mu(T_S)$ is for Uniform distribution, and the best value of $\sigma(T_S)$ is for Brownian distribution. So the combination of Levy distribution for level 1 search and Brownian distribution for level 2 search is the most efficient among all in confirming fixed targets.
\begin{table}[h]
\centering
\caption{Effect of probability distribution on confirmation of fixed targets ($n_i$=10 and $n_u$=12) }
\label{simtable4}
\begin{tabular}{ l c  c  c  c }
\hline
 Level 1 & Level 2   &  $\mu(T_{nc})\pm\sigma(T_{nc})$    &  $\mu(T_S)\pm\sigma(T_S)$  \\ 
 &&& (seconds)\\
\hline
 US  & US &  4.231$\pm$1.416  &  4338$\pm$2024  \\
 US  & BS   &  4.341$\pm$1.409  &  4521$\pm$2113   \\
 US  & LS       &  4.130$\pm$1.397  &  5166$\pm$2561  \\ \hline
 BS & US    &  4.228$\pm$1.358  &  4451$\pm$2208   \\
 BS & BS   &  4.282$\pm$1.351  &  4662$\pm$2365 \\
 BS & LS       &  4.184$\pm$1.389  &  5317$\pm$3186  \\\hline
 LS     & LS       &  4.687$\pm$1.449  &  4535$\pm$2323  \\ 
LS     & US    &  4.617$\pm$1.441  &  3888$\pm$1884  \\ 
 LS     & BS   &  4.658$\pm$1.396  &  4001$\pm$1834  \\
\hline
\end{tabular}
\end{table}

The effect of using a different probability distribution for level 1 search and level 2 search of MDM framework for confirmation of 10 pop-up targets is shown in Table \ref{simtable5}. The best search performance in terms of $\mu(T_{nc})$ is given by combining uniform distribution for level 1 search and Brownian distribution for level 2 search. The best value of $\sigma(T_{nc})$ is obtained by combining Brownian and Levy distributions. The best value of $\mu(T_S)$ and $\sigma(T_{s})$ is obtained by Uniform distribution for both level 1 search and level 2 search. Thus, Uniform distribution is the best for level 1 search and for level 2 search in confirming sudden pop-up targets.

\begin{table}[h]
\centering
\caption{Effect of probability distribution on confirmation of  pop-up targets ($n_i$=0, max($n_p$)=10 and $n_u$=12)}
\label{simtable5}
\begin{tabular}{ l c  c  c  c }
\hline
 Level 1 & Level 2   &  $\mu(T_{nc})\pm\sigma(T_{nc})$    &  $\mu(T_S)\pm\sigma(T_S)$  \\ 
 &&& (seconds)\\
\hline
 US  & US    &  5.106$\pm$1.348  &  4996$\pm$1800  \\
 US  & BS   &  5.151$\pm$1.388  &  5066$\pm$1829  \\
 US  & LS       &  5.046$\pm$1.318  &  5185$\pm$1956  \\ \hline
 BS & US    &  5.085$\pm$1.327  &  5276$\pm$2116  \\
 BS & BS   &  5.081$\pm$1.337  &  5285$\pm$2067  \\
 BS & LS       &  5.031$\pm$1.306  &  5324$\pm$2157  \\ \hline
 LS     & LS       &  5.094$\pm$1.339  &  5269$\pm$2173  \\ 
 LS     & US    &  5.082$\pm$1.325  &  5106$\pm$2013  \\ 
 LS     & BS   &  5.085$\pm$1.332  &  5293$\pm$2127  \\
\hline
\end{tabular}
\end{table}

The effect of using a different probability distribution for level 1 search and level 2 search of MDM framework for confirmation of 10 dynamic targets is shown in Table \ref{Dsimtable6}. The best average search performance and completeness of the search mission are obtained by combining the Levy distribution for the level 1 search and the Brownian distribution for the level 2 search. For further analysis, a combination of LS (level 1) and BS (level 2) is used for the case of fixed targets and dynamic targets and the US (for both level 1 and level 2) is used for the case of sudden pop-up targets with step lengths $\delta_1$=20$\%$ of L and $\delta_2=r_1$ respectively.

\begin{table}[h]
\centering
\caption{Effect of probability distribution on confirmation of  dynamic targets ($n_d$=10 and $n_u$=12)}
\label{Dsimtable6}
\begin{tabular}{ l c  c  c  c }
\hline
 Level 1 & Level 2   &  $\mu(T_{nc})\pm\sigma(T_{nc})$    &  $\mu(T_S)\pm\sigma(T_S)$  \\ 
 &&& (seconds)\\
\hline
 US & US & 4.383$\pm$1.499 & 2678$\pm$1047 \\
 US & BS & 4.319$\pm$1.516 & 2767$\pm$1019 \\
 US & LS & 4.337$\pm$1.484 & 2759$\pm$1082 \\ \hline
 BS & US & 4.333$\pm$1.566 & 2674$\pm$1061 \\
 BS & BS & 4.346$\pm$1.497 & 2686$\pm$1086 \\
 BS & LS & 4.409$\pm$1.487 & 2671$\pm$1039 \\ \hline
 LS & LS & 4.351$\pm$1.534 & 2674$\pm$1029 \\ 
 LS & US & 4.419$\pm$1.542 & 2630$\pm$1043 \\ 
 LS & BS & 4.424$\pm$1.499 & 2382$\pm$893  \\
\hline
\end{tabular}
\end{table}

\subsection{Performance comparison with self-cognitive search}
To evaluate the benefits of the MDM framework over self-cognitive based single sensor search (SSS), four stochastic search approaches are considered in the UAVs search mission to locate fixed/pop-up/dynamic targets. This includes three SSS techniques, (i) the US with samples drawn from the uniform distribution; (ii) BS with samples drawn from the normal distribution; (iii) LS with samples drawn from the Levy distribution; and (iv) the proposed MDM framework that adaptively switches between the LS and the BS for the case of initially fixed targets and dynamic targets and the US (for both levels of search) for the case of sudden pop-up targets. Here, 1000 Monte-Carlo simulations are carried out with two performance criteria for searching initially fixed and sudden pop-up targets. The step length for US, BS, and LS techniques are taken as 20$\%$ of L with only a confirmation sensor employed for search. The average improvement in search efficiency (AISE) of MDM framework is computed as,
\begin{equation}
    AISE=\frac{\mu(T_{nc})_{MDM}-\mu(T_{nc})_{SSS}}{\mu(T_{nc})_{SSS}} \times 100 \%
\end{equation}
where $\mu(T_{nc})_{MDM}$ and $\mu(T_{nc})_{SSS}$ are the average number of targets confirmed for the MDM and SSS respectively. The average improvement in search completeness (AISC) of MDM framework is computed as,
\begin{equation}
    AISC=\frac{\mu(T_{S})_{SSS}-\mu(T_{S})_{MDM}}{\mu(T_{S})_{SSS}} \times 100 \%
\end{equation}
where $\mu(T_{S})_{SSS}$ and $\mu(T_{S})_{MDM}$ are the average time taken to confirm 10 targets for SSS and MDM respectively.  

The results of 1000 Monte-Carlo simulations for confirming initially fixed 10 targets with 12 UAVs are shown in Table \ref{simtable1}. From this Table, it is clear that the MDM search performs the best among all the four search techniques for all cases (fixed, pop-up and dynamic targets).
\begin{table}[h]
\centering
\caption{Comparison between MDM and self-cognitive based search for confirmation of 10 targets with 12 UAVs}
\label{simtable1}
\begin{tabular}{  c  c  c  c c}
\hline
 Search        &  $\mu(T_{nc})\pm\sigma(T_{nc})$      &  AISE& $\mu(T_S)\pm\sigma(T_S)$ & AISC   \\ 
       technique    &   &   ($\%$)             &    (seconds) &  ($\%$)         \\ 
\hline
Fixed targets &&&&\\
\hline
 US &  4.086$\pm$1.472  & 12.3 & 7879$\pm$4811  & 49.2\\
 BS      &  3.985$\pm$1.515  & 14.4 & 9426$\pm$6391 & 57.6\\
 LS          &  3.787$\pm$1.457  & 18.7 &  9308$\pm$6021 & 57.0\\
 MDM (LS+BS)         &  4.658$\pm$1.396  & - & 4001$\pm$1834 & -\\
\hline
Pop-up targets &&   &&\\
\hline
 US &  4.103$\pm$1.358 & 24.5  & 9316$\pm$4866 & 46.4\\
 BS &  4.082$\pm$1.361 & 25.1  & 10704$\pm$6328 & 53.3\\
 LS &  4.047$\pm$1.416 & 26.2  & 10802$\pm$6388 & 53.8\\
 MDM (US+US) & 5.106$\pm$1.348 & - & 4996$\pm$1800 & - \\
\hline
Dynamic targets &&   &&\\
\hline
US &  3.859$\pm$1.465 & 14.6 & 3807$\pm$1602 & 37.4\\
BS &  3.811$\pm$1.457 & 16.1 & 3689$\pm$1453 & 35.4\\
LS &  3.692$\pm$1.473 & 19.8 & 3729$\pm$1569 & 36.1\\
MDM (LS+BS) &  4.424$\pm$1.499 & - & 2382$\pm$893 & -\\
 \hline
\end{tabular}
\end{table}

\begin{figure}[h!]
    \centering
    \includegraphics[height=6cm, width=9cm]{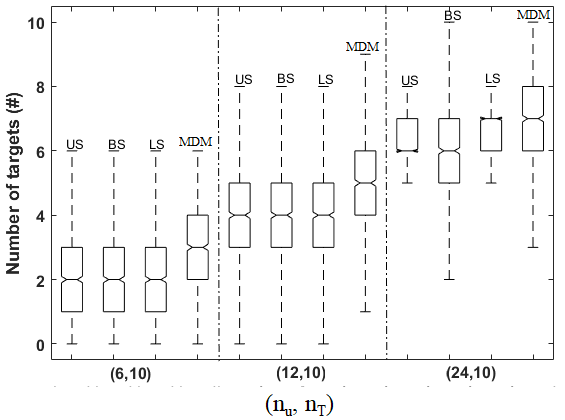}
    \caption{Efficiency of search techniques for different number of UAVs for the case of fixed targets}
    \label{scalefiga}
\end{figure}

\begin{figure}[h!]
    \centering
    \includegraphics[height=6cm, width=9cm]{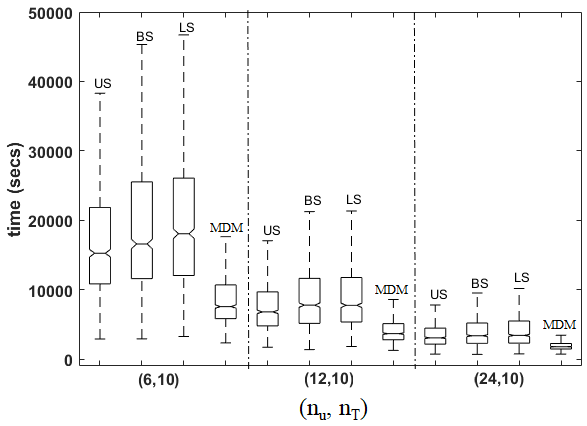}
    \caption{Completeness of search techniques for different number of UAVs for the case of fixed targets}
    \label{scalefigb}
\end{figure}

\noindent \textit{$B_1$: Effect of the number of UAVs}\\
The search technique is simulated for a different number of UAVs ($n_u$) involved in the search mission with 10 initially fixed targets ($n_i$). Monte-Carlo runs are conducted for $n_u=\{6,\,\,12,\,\,24\}$ with the target position kept the same for all the search techniques. The efficiency of all four search techniques is compared in Fig. \ref{scalefiga}. For the case of 6, 12 and 24 UAVs, the MDM framework (LS+BS) has the highest value for the maximum and the minimum number of targets confirmed. For all three cases, the MDM framework (LS+BS) has the highest average number of targets confirmed. Completeness of the search techniques for the case of initially fixed targets is shown in Fig. \ref{scalefigb}. The worst-case time taken by the MDM framework (LS+BS) to confirm all the targets is 50$\%$ less than the worst-case time taken by other search techniques.

\begin{figure}[h!]
    \centering
    \includegraphics[height=6cm, width=9cm]{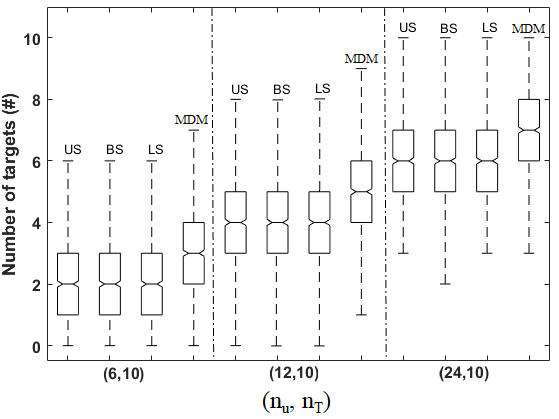}
    \caption{Efficiency of search techniques for different number of UAVs for the case of pop-up targets}
    \label{scalefigc}
\end{figure}

\begin{figure}[h!]
    \centering
    \includegraphics[height=6cm, width=9cm]{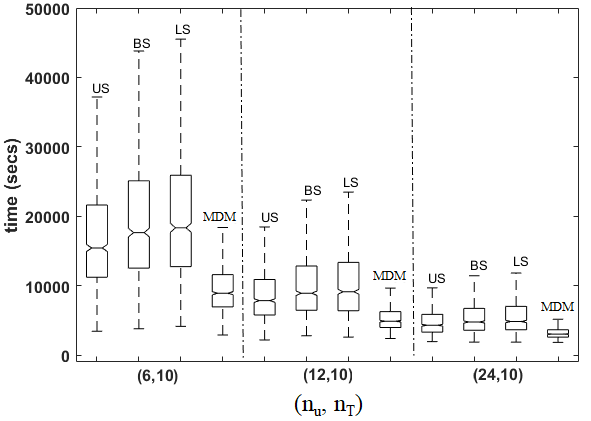}
    \caption{Completeness of search techniques for different number of UAVs for the case of pop-up targets}
    \label{scalefigd}
\end{figure}

\begin{figure}[h!]
    \centering
    \includegraphics[height=6cm, width=9cm]{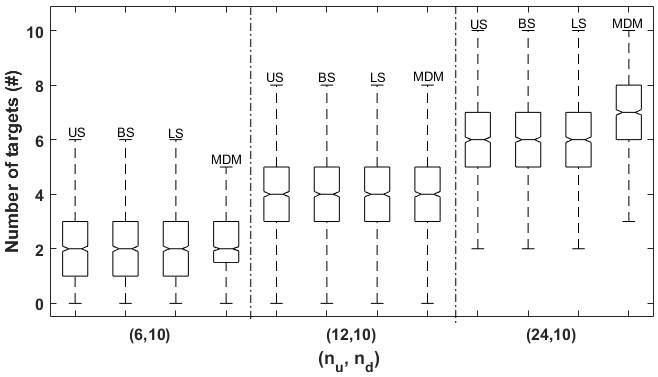}
    \caption{Efficiency of search techniques for different number of UAVs for the case of dynamic targets}
    \label{Dscalefigc}
\end{figure}

\begin{figure}[h!]
    \centering
    \includegraphics[height=6cm, width=9cm]{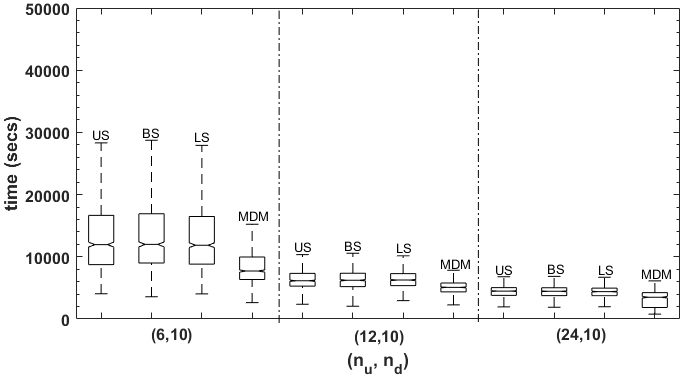}
    \caption{Completeness of search techniques for different number of UAVs for the case of dynamic targets}
    \label{Dscalefigd}
\end{figure}

Fig. \ref{scalefigc} shows the box plot of the search with sudden pop-up targets. For the case of 6 and 12 UAVs, the MDM search (US+US) is able to confirm a maximum of 7 and 9 targets, respectively, whereas other search techniques could confirm only 6 and 8 targets. For all three cases, the average number of targets searched by the MDM search (US+US) is higher than other search techniques. Fig. \ref{scalefigd} shows the box plot for search performance with varying the number of UAVs where this is useful to analyze the time taken to search 10 targets. From Fig. \ref{scalefigd}, we can observe that for ($n_u$, max($n_p$)) to be {(6,10), (12,10), (24,10)} our proposed MDM search (US+US) has the lowest best case and worst case search time required in comparison with the US, BS and LS. Also, the first and second quartile range is lesser for the MDM framework (US+US) than US, BS, and LS. For a fixed area of operation with a fixed number of targets, the advantage of the MDM framework is higher as the number of UAVs increases. Figs. \ref{Dscalefigc} and \ref{Dscalefigd} show the box plot of the efficiency and completeness of search techniques for dynamic targets. In Fig. \ref{Dscalefigc}, we can observe that the proposed search method is able to achieve better search efficiency in the case of 24 UAVs and 10 targets, whereas, for the other two cases, the search efficiency remains the same for all the 4 methods. However, for the case of search completeness given in Fig. \ref{Dscalefigd}, we can observe that the MDM search (LS+BS) has the lowest best-case time required to search all the targets in comparison with the US, BS and LS.  

\noindent\textit{$B_2$: Effect of false target detection}\\
The use of low-accuracy sensors for initial target detection might result in the unfavorable detection of false targets. Out of 10 targets, 3 targets are classified as false targets for the Monte-Carlo runs. The results for 1000 Monte-Carlo runs are shown in Table \ref{simtablef1}. The proposed MDM search is at least 21$\%$ better than SSS for the case of AISE and 45.2$\%$ for the case of AISC. For the case of pop-up targets, an improvement of at least 22.3$\%$ and 31.9$\%$ is observed. For the case of dynamic targets, an improvement of at least 18.1$\%$ and 23.1$\%$ is observed in AISE and AISC respectively.

\begin{table}[h]
\centering
\caption{Comparison between SSS and MDM search for confirmation of 10 targets (7 real and 3 false) with 12 UAVs }
\label{simtablef1}
\begin{tabular}{  c  c  c  c c}
\hline
 Search        &  $\mu(T_{nc})\pm\sigma(T_{nc})$      &  AISE& $\mu(T_S)\pm\sigma(T_S)$ & AISC   \\ 
       technique    &   &   ($\%$)             &    (seconds) &  ($\%$)         \\ 
\hline
Fixed targets &&&&\\
\hline
 US &  2.915$\pm$1.320  & 21.0 & 7943$\pm$5018  & 45.2\\
 BS      &  2.866$\pm$1.298  & 23.1 & 9242$\pm$6427 & 52.9\\
 LS          &  2.775$\pm$1.259  & 27.1 &  9416$\pm$6676 & 53.8\\
 MDM (LS+BS)         &  3.528$\pm$1.246  & - & 4352$\pm$2705 & -\\
\hline
Pop-up targets &&&&\\
\hline
 US &  2.887$\pm$1.208  & 22.3 & 8634$\pm$4592  & 31.9\\
 BS      &  2.747$\pm$1.159  & 28.5 & 10457$\pm$6104 & 43.7\\
 LS          &  2.766$\pm$1.146  & 27.6 &  10116$\pm$5884 & 41.8\\
 MDM (US+US)         &  3.530$\pm$1.154  & - & 5883$\pm$2868 & -\\
 \hline
Dynamic targets &&&&\\
\hline
US & 2.637$\pm$1.278 & 19.5 & 6995$\pm$2954 & 23.1 \\
BS & 2.578$\pm$1.266 & 22.2 & 7226$\pm$3121 & 25.5 \\
LS & 2.669$\pm$1.270 & 18.1 & 7111$\pm$2841 & 24.3 \\
MDM (LS+BS) & 3.151$\pm$1.273 & - & 5381$\pm$1717 & -\\
\hline
\end{tabular}
\end{table}

\subsection{Performance comparison with social-cognitive search}
The proposed MDM search is compared with the well-known multi-UAV search that employs both self and social-cognitive elements, namely, Standard Particle Swarm Optimization (PSO) \cite{pso}, Standard PSO (SPSO) \cite {ppso18} and Adaptive Robotic PSO (ARPSO) \cite{ARPSO}. The key assumption for social cognition is that UAVs have persistent communication throughout the mission. Practically this assumption might not be feasible, whereas the proposed MDM framework does not require communication among UAVs. In all the above methods, the UAVs are carrying a sensor that can generate a signal with strength proportional to the distance from the target. A target is confirmed when the distance to the target is less than $r_2$, i.e. the sensing radius of the confirmation sensor used in the MDM search. 

The results of the Monte-Carlo simulation for the confirmation of 10 fixed and also pop-up targets using 12 UAVs for the search mission are shown in Table \ref{simtablepso1}. The MDM search with an additional detection sensor performs better compared with the social-cognitive based multi-UAV search methods. For the case of targets with fixed location and dynamic, MDM performs better than PSO, SPSO and ARPSO. For the case of sudden pop-up targets, the proposed MDM search outperforms social-cognitive multi-UAV search methods by a margin of at least 73.6$\%$. The lower performance of the social-cognitive multi-UAV search methods for pop-up targets is because of the skewness of the spatial distribution of UAVs towards the region where the targets have already been detected. This results in a less efficient social-cognition component for a new target that suddenly pop-up. In addition, social-cognitive multi-UAV search methods can explore and later has better exploitation capability, whereas the MDM framework will keep switching between exploration and exploitation. Hence, the MDM search pop-up target is better than the social-cognitive multi-UAV search methods. From the above analysis, the performance of the MDM search without inter-UAV communication is comparable to most of the social-cognitive multi-UAV search methods that employ inter-UAV communication.

\begin{table}[h]
\centering
\caption{Comparison between the MDM framework and social-cognition based search methods for 10 targets with 12 UAVs}
\label{simtablepso1}
\begin{tabular}{l  c  c c c}
\hline
Case & Search   &  $\mu(T_{nc})\pm\sigma(T_{nc})$      &  AISE($\%$) & Commun- \\ 
     &   methods &   & & ication\\
\hline
      & PSO   & 4.281$\pm$1.233 & 8.8 & Yes \\

Fixed & SPSO & 4.422$\pm$1.291 & 5.3 & Yes\\

targets & ARPSO & 4.319$\pm$1.401 & 7.8 & Yes\\

 & MDM (LS+BS) & 4.658$\pm$1.396 & - & No\\

\hline
 & PSO  & 2.94$\pm$1.29 & 73.6 & Yes\\

Pop-up & SPSO & 2.88$\pm$1.36 & 77.2 & Yes\\

targets & ARPSO & 2.685$\pm$1.487 & 90.1 & Yes\\

 & MDM (US+US) & 5.106$\pm$1.348 & - & No\\
\hline
 & PSO & 4.244$\pm$1.223 & 4.24 & Yes\\ 

Dynamic & SPSO & 4.370$\pm$1.271 & 1.24 & Yes\\ 

targets & ARPSO & 4.285$\pm$1.396 & 3.24 & Yes\\

 & MDM (LS+BS) &  4.424$\pm$1.499 & - & No\\
\hline
\end{tabular}
\end{table}

\section{Conclusions}
In this paper, a new metacognition decision framework based on decentralized multi-UAV search has been implemented. Each UAV performs the MDM framework search to confirm fixed, sudden pop-up and dynamic targets. The proposed search method without communication among UAVs is appropriate for searching fixed, pop-up (evolves over time) and dynamic targets distributed over a larger region. The paper provides a generic framework to utilize information from multiple sensors for target search. The theoretical analysis indicates that the probability of confirming a target for the MDM search is greater than or equal to the conventional cognition-based multi-UAV search for a given search time duration.

The proposed MDM search is simulated for two sensors. A combination of LS for the first level and BS for the second level is found to be more effective for searching fixed and dynamic targets. For sudden pop-up targets, applying the US for both levels is found to give the best results. The Monte-Carlo simulation shows that the average efficiency of the MDM search is at least 12$\%$ better than other single sensor-based stochastic search techniques for fixed targets. For the case of pop-up and dynamic targets search, the MDM framework is at least 24$\%$ and 14$\%$ respectively more efficient on average than other stochastic search techniques. The completeness of the MDM framework is at least 49$\%$, 46$\%$ and 35$\%$ better than other single sensor-based stochastic (SSS) search techniques for the case of fixed, sudden pop-up and dynamic targets, respectively. For a given area of operation, the performance of the proposed search algorithm increases with an increase in the number of UAVs in comparison with other methods. The performance of the proposed  MDM framework is also comparable to the well-known social-cognition based search techniques like PSO, SPSO and ARPSO. The Monte-Carlo simulation indicates that the MDM framework is more efficient in searching targets distributed in unknown environments when compared to other self-cognitive SSS search techniques and also the PSO, SPSO, and ARPSO-based search needs communication. Currently, this framework is tested for UAVs carrying two sensors (single detection and confirmation). This can be extended to a higher number of detection sensors, which will be explored in the future.

\begin{IEEEbiography}[{\includegraphics[width=1in,height=1.25in,clip,keepaspectratio]{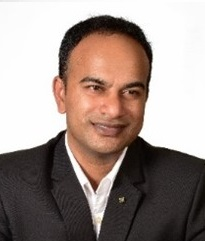}}]{J. Senthilnath} (Senior Member, IEEE) received the Ph.D. degree in aerospace engineering from the Indian Institute of Science, Bengaluru, India, in 2014. He is a Senior Scientist with the Institute for Infocomm Research, Agency for Science, Technology, and Research (A*STAR), Singapore. His research interests include generative models, reinforcement learning, online learning, and optimization in multiagent systems, materials informatics, failure analysis, and remote sensing. 
\end{IEEEbiography}

\begin{IEEEbiography}[{\includegraphics[width=1in,height=1.25in,clip,keepaspectratio]{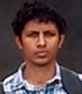}}]{K. Harikumar} received the Ph.D. degree in aerospace engineering from the Indian Institute of Science, Bengaluru, India, in 2015. He is working as an Assistant Professor with the Robotics Research Centre, International Institute of Information Technology, Hyderabad, India. His research interests are applications of control theory to unmanned systems and flight dynamics.
\end{IEEEbiography}

\begin{IEEEbiography}[{\includegraphics[width=1in,height=1.25in,clip,keepaspectratio]{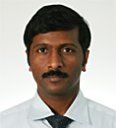}}]{Suresh Sundaram} (Senior Member, IEEE) received the Ph.D. degree in aerospace engineering from the Indian Institute of Science, Bengaluru, India, in 2005. He is currently an Associate Professor with the Department of Aerospace Engineering, Indian Institute of Science. From 2010 to 2018, he has been an Associate Professor with the School of Computer Science and Engineering, Nanyang Technological University. His research interests include flight control, unmanned aerial vehicle design, machine learning, optimization, and computer vision.
\end{IEEEbiography}

\end{document}